\documentclass[runningheads]{llncs}
\usepackage[T1]{fontenc}
\usepackage{multirow}
\usepackage{makecell}
\usepackage{graphicx}
\usepackage{amsmath}
\usepackage{amssymb}
\usepackage{color}
\usepackage{xfrac}
\usepackage{wrapfig}
\usepackage{cite}
\usepackage[misc]{ifsym}
\usepackage[colorlinks,
            urlcolor=red,
            hyperfootnotes = True,
            citecolor=green]{hyperref}
\usepackage{bm}
\def\A{\bm{A}}

\def\C{\bm{C}}

\def\M{\bm{M}}

\def\X{\bm{X}}
\def\Y{\bm{Y}}

\def\I{\bm{I}}

\def\mM{\mathcal{M}}

\def\mC{\mathcal{C}}

\begin{document}
\title{Orientation-Shared Convolution Representation for CT Metal Artifact Learning}
\titlerunning{Orientation-Shared Convolutional Network}
%

\author{Hong Wang\inst{1} \and
Qi Xie\inst{2}\textsuperscript{(\Letter)}\and
Yuexiang Li\inst{1}\textsuperscript{(\Letter)}\and
Yawen Huang\inst{1}\and \\
Deyu Meng\inst{2,3,4}\and
Yefeng Zheng\inst{1}}

\authorrunning{H. Wang et al.}
%
%

\institute{Tencent Jarvis Lab, Shenzhen, P.R. China\\
\email{\{hazelhwang,vicyxli,yawenhuang,yefengzheng\}@tencent.com}\and
Xi'an Jiaotong University, Xi'an, Shaan'xi, P.R. China \\
\email{\{xie.qi,dymeng\}@mail.xjtu.edu.cn}\and
Peng Cheng Laboratory, Shenzhen, China\and
Macau University of Science and Technology, Taipa, Macau
}

\maketitle              
\begin{abstract}
During X-ray computed tomography (CT) scanning, metallic implants carrying with patients often lead to adverse artifacts in the captured CT images and then impair the clinical treatment. Against this metal artifact reduction (MAR) task, the existing deep-learning-based methods have gained promising reconstruction performance. Nevertheless, there is still some room for further improvement of MAR performance and generalization ability, since some important prior knowledge underlying this specific task has not been fully exploited. Hereby, in this paper, we carefully analyze the characteristics of metal artifacts and propose an orientation-shared convolution representation strategy to adapt the physical prior structures of artifacts, \emph{i.e.}, rotationally symmetrical streaking patterns. The proposed method rationally adopts Fourier-series-expansion-based filter parametrization in artifact modeling, which can better separate artifacts from anatomical tissues and boost the model generalizability. Comprehensive experiments executed on synthesized and clinical datasets show the superiority of our method in detail preservation beyond the current representative MAR methods. Code will be available at \url{https://github.com/hongwang01/OSCNet}.
\end{abstract}

\section{Introduction}
During the computed tomography (CT) imaging process, metallic implants within patients would severely attenuate X-rays and lead to the missing X-ray projections. Accordingly, the captured CT images often present streaking and shading artifacts~\cite{lin2019dudonet,liao2019adn,wang2021dicdnet}, which negatively affect the clinical treatment.

Against this metal artifact reduction (MAR) task, many approaches have been proposed in the past few years. Traditional methods replaced the metal-corrupted region in sinogram domain with surrogates, which were estimated via different manners, \emph{e.g.,} linear interpolation (LI)~\cite{kalender1987reduction} and normalized MAR~\cite{meyer2010normalized}. Recently, driven by deep learning (DL) techniques, some researchers exploited the powerful fitting capability of deep neural networks to directly reconstruct clean sinogram~\cite{park2018ct,gjesteby2017deep,zhang2018convolutional,liao2019generative,ghani2019fast}. However, the estimated values in sinogram domain are not always consistent with the physical imaging geometry, which leads to the secondary artifacts in the reconstructed CT images. Furthermore, it is difficult to collect the sinogram data in realistic applications \cite{liao2019adn}. These limitations constrain the generality and performance of sinogram-based MAR approaches.

To loosen the requirement of sinogram, researchers proposed to train deep networks based on only CT images for the recovery of clean CT images~\cite{huang2018metal,gjesteby2018deep,wang2018conditional}. Recently, some works simultaneously exploited sinogram and CT image~\cite{wang2021indudonet, wang2021indudonet+,lin2019dudonet,yu2020deep,lyu2020dudonet++,wang2021dualclean,zhou2022dudodr} to further boost the MAR performance. Nevertheless, the involved image domain-based network module of most methods is a general structure for image restoration tasks, which has not sufficiently encoded useful priors underlying this specific MAR task and made the network hardly interpretable.

\begin{figure}[t]
  \begin{center}
   \vspace{-4mm}
     \includegraphics[width=0.92\linewidth]{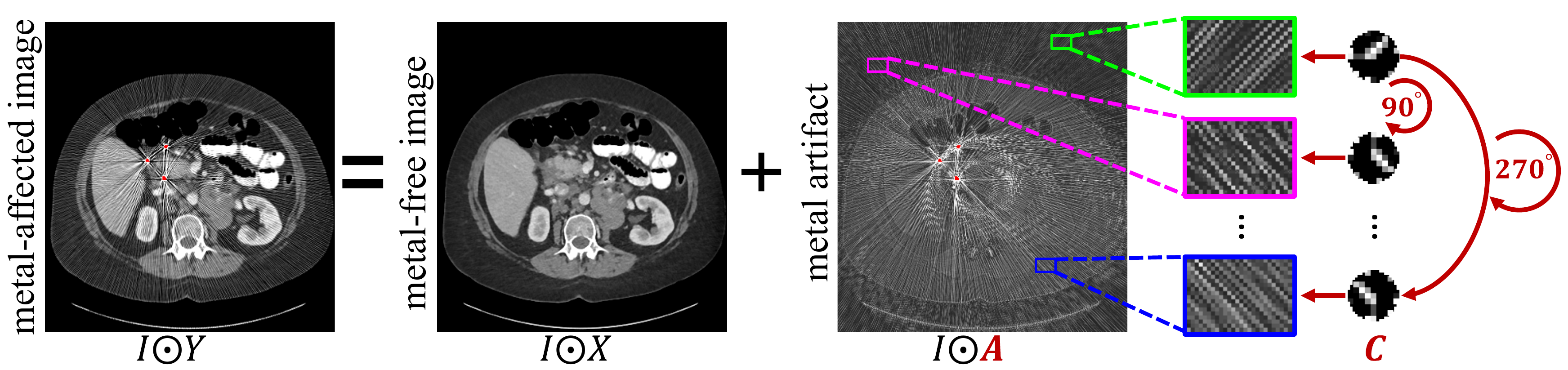}
  \end{center}
  \vspace{-7mm}
     \caption{Metal artifact $\A$ presents clear rotationally streaking structures, which can be represented by rotating each convolution filter $\C$ to multiple angles (\emph{i.e.}, orientation-shared). Here $\Y$, $\X$, and $\I$ represent the metal-affected CT image, ground truth image, and non-metal region, respectively. Red pixels in $\Y$ and $\A$ are metallic implants.}
  \label{figintro}
    \vspace{-4mm}
\end{figure}

To alleviate these issues, Wang \emph{et al.}~\cite{wang2021dicdnet} proposed an interpretable MAR network named as DICDNet, which is integrated with the non-local repetitive streaking priors of metal artifacts $\A$ via a convolutional dictionary model as:
\vspace{-2mm}
\begin{equation}\label{eqa}
    \A = \sum_{s=1}^{S} \C_{s}  \otimes \M_{s},
    \vspace{-2mm}
\end{equation}
where the filter $\C_{s}$ and the feature map $\M_{s}$ represent the local streaking patterns and the locations of artifacts, respectively; $\otimes$ denotes 2D convolution operator.

However, there is still some important and insightful prior knowledge not encoded by Eq.~\eqref{eqa}. Specifically, one of the most intrinsic priors is that along every rotation direction, the caused artifacts have similar streaking patterns, as shown in Fig.~\ref{figintro}. Such inherent \textbf{rotationally symmetrical streaking (RSS)} prior structures are often caused by the rotationally scanning-based CT physical imaging process~\cite{zhang2018convolutional,liao2019adn}, and appear in diverse kinds of CT imaging-related artifacts. Besides, it should be noted that the RSS structure is specifically possessed by metal artifact $\A$ rather than the to-be-estimated clean CT image $\X$. This fact implies that it would be useful and valuable to encode and utilize such RSS prior structure so as to distinguish metal artifacts from clean CT images.

To encode such inherent RSS prior structure, we seek to construct a novel convolutional dictionary model that shares filters among different angles. In other words, we aim to represent the metal artifact $\A$ along multiple directions by rotating each convolution filter $\C$ to multiple angles, as shown in Fig.~\ref{figintro}.  In this manner, we only need to learn the filter $\C$ along one angle, which can represent local patterns at different angles but with similar structures (\emph{i.e.}, orientation-shared). Naturally, this helps reduce the number of learnable filters and improve the flexibility of network representation. However, the commonly-used discrete convolution filters can hardly be rotated to different angles with high precision. Thus, we adopt the filter parametrization method~\cite{xie2021fourier} and firstly propose an orientation-shared convolution representation model, which finely encodes the RSS prior of artifacts. Based on the proposed model, we construct an optimization-inspired network for artifact learning. Compared to current state-of-the-art (SOTA) MAR methods, our method can better reduce artifacts while more faithfully preserving tissues with fewer model parameters for representing artifacts, which are finely verified by experiments on synthetic and clinical data.

\vspace{-2mm}
\section{Preliminary Knowledge} 
\vspace{0mm}
Filter parametrization is an important strategy to achieve parameter sharing among convolution filters at different rotation angles. Since 2D discrete Fourier transform can
be equivalently expressed as Fourier series expansion, it is natural to construct a Fourier-series-expansion-based filter parametrization method~\cite{xie2021fourier}. Specifically, to represent a discrete filter $\bm{C} \in \mathbb{R}^{p\times p}$, we can discretrize a 2D function $\varphi\left(x\right)$ by uniformly sampling on the area within $[\sfrac{-(p-1)h}{2}, \sfrac{(p-1)h}{2}]^2$, where $h$ is the mesh size of images, and $\varphi\left(x\right)$ can be expressed as~\cite{cooley1969fast}:
\vspace{-2mm}
\begin{equation}\label{eq1}
    \varphi\left(x\right)=  \sum_{q=0}^{p-1} \sum_{t=0}^{p-1} a_{qt} \varphi_{qt}^{c}\left(x\right) +b_{qt}\varphi_{qt}^{s}\left(x\right),
    \vspace{-2mm}
\end{equation}
where $x=[x_{i},x_{j}]^{T}$ is 2D spatial coordinate; $a_{qt}$ and $b_{qt}$ are expansion coefficients; $\varphi_{qt}^{c}\left(x\right)$ and $\varphi_{qt}^{s}\left(x\right)$ are 2D basis functions. In conventional Fourier series, one can select the basis functions as:
\vspace{-1mm}
\begin{equation}\label{eq2}
\small
\varphi_{qt}^{c}\left(x\right)=\Omega \left(x\right)\text{cos}\left(\frac{2\pi}{ph}\left[q,t\right]
\cdot \begin{bmatrix}  
 x_{i}\\
x_{j}
\end{bmatrix}\right),~~\varphi_{qt}^{s}\left(x\right)=\Omega \left(x\right)\text{sin}\left(\frac{2\pi}{ph}\left[q,t\right]
\cdot \begin{bmatrix}
 x_{i}\\
x_{j}
\end{bmatrix}\right),
\vspace{-1mm}
\end{equation}
where $\Omega \left(x\right)\ge0$ is a radial mask function and $\Omega \left(x\right)=0$ if $\left \|  x\right \|\ge (\sfrac{(p+1)}{2})h$. This facilitates the rotation operation~\cite{xie2021fourier}.


Unfortunately, the Fourier bases~\eqref{eq2} always present aliasing effect when $\varphi\left(x\right)$ is rotated. To alleviate this issue, Eq.~\eqref{eq2} is revised in~\cite{xie2021fourier} as:
\begin{equation}\label{eq3}
\small
\varphi_{qt}^{c}\!\left(x\right)\!=\!\Omega \left(x\right)\!\text{cos}\!\left(\!\frac{2\pi}{ph}\!\left[\mathcal{I}_{p}\left(q\right),\mathcal{I}_{p}\left(t\right)\right]
\cdot \begin{bmatrix}  
 x_{i}\\
x_{j}
\end{bmatrix}\right),\varphi_{qt}^{s}\!\left(x\right)\!=\!\Omega \left(x\right)\!\text{sin}\!\left(\!\frac{2\pi}{ph}\left[\mathcal{I}_{p}\left(q\right),\mathcal{I}_{p}\left(t\right)\right]
\cdot \begin{bmatrix}
 x_{i}\\
x_{j}
\end{bmatrix}\!\right),
\end{equation}
where $\text{if}~ y\le \sfrac{p}{2},~\mathcal{I}_{p}\left(y\right)=y$;~otherwise, $\mathcal{I}_{p}\left(y\right)= y-p.$

\vspace{-2mm}
\section{Orientation-Shared Convolution Model for MAR}
\vspace{-1mm}
Clearly, the metals generally have extremely higher CT values than normal tissues. From Fig.~\ref{figintro}, it is easily observed that for the MAR task, our goal is to reconstruct the clean tissues in the non-metal region from a given metal-corrupted CT image $\Y\in\mathbb{R}^{H\times W}$, without paying attention to the learning in the metal region. Similar to~\cite{wang2021dicdnet}, we define the decomposition model as:
\begin{equation}\label{eq4}
\I\odot\Y=\I\odot\X +\I\odot\A,
\end{equation}
where $\X$ and $\A$ are the unknown metal-free CT image and the artifact layer, respectively; $\I$ is a binary non-metal mask; $\odot$ denotes point-wise multiplication.

Clearly, estimating $\X$ from $\Y$ is an ill-posed inverse problem, and rationally utilizing prior knowledge can finely help constrain the solution space and better reconstruct clean images.
To this end, we carefully analyze the physical priors of artifacts and accordingly formulate them as an explicit model. The prior model is then embedded into deep networks to regularize the extraction of artifacts.

\vspace{2mm}
\noindent \textbf{Metal Artifact Modeling with Filter Parametrization.}
Specifically, since the CT imaging is generally performed in a rotationally scanning manner, metal artifacts generally present scattered streaking structures~\cite{liao2019adn,lin2019dudonet}.  From Fig.~\ref{figintro}, we can easily find that along every rotation angle, artifacts share similar streaking patterns. To encode such prior structures, we can rotate convolution filters to multiple angles so as to represent artifacts in different directions. To this end, we firstly propose a filter parametrization based convolutional coding model as:
\vspace{-2mm}
\begin{equation}\label{eq7}
\A =  \sum_{l=1}^{L} \sum_{k=1}^{K} \C_{k}(\theta_{l})  \otimes \M_{lk},
\vspace{-2mm}
\end{equation}
where $L$ is the number of rotation angles; $\theta_{l}=\sfrac{2\pi\left(l-1\right)}{L}$ is the $l$-th rotation angle; $K$ is the number of convolution filters at each angle; $\C_{k}(\theta_{l})\in \mathbb{R}^{p \times p}$ is the $k$-th parametrized filter at angle $\theta_{l}$, and it represents the streaking and rotated prior patterns of artifacts; $\M_{lk}$ is feature map reflecting the locations of artifacts.

From Eq.~\eqref{eq7}, one can easily find that metal artifacts in different orientations share the same set of filters as $\C_{k}(\theta_{l}) (k=1,\ldots,K)$ under orientation-variance freedom. This finely encodes the fact that metal artifacts in different directions share similar patterns. Compared to the conventional convolutional dictionary model~\eqref{eqa} in~\cite{wang2021dicdnet}, our proposed model~\eqref{eq7} has two main merits: 1) Eq.~\eqref{eq7} should be more accurate for representing artifacts than model~\eqref{eqa}, since it can ensure the RSS prior of artifacts (see Fig.~\ref{figverf} below); 2) The number of the learnable filters can be evidently reduced, which would benefit the learning of $\C_{k}(\theta_{l})$.



Specifically, for modeling $\C_{k}(\theta_{l})$, it is necessary to adopt a filter parametrization method with high representation accuracy. The work~\cite{xie2021fourier} has shown that the basis function in Eq.~\eqref{eq3} can represent arbitrary filters with arbitrary angles. Based on this, we can parametrize every element in $\C_{k}(\theta_{l})$ as:
\vspace{-2mm}
\begin{equation}\label{eq8}
[\C_{k}(\theta_{l})]_{ij}={\varphi}_{k}\left(T_{\theta_{l}}x_{ij}\right) =\sum_{q=0}^{p-1} \sum_{t=0}^{p-1} a_{qtk} {\varphi}_{qt}^{c}\left(T_{\theta_{l}}x_{ij}\right) +b_{qtk}{\varphi}_{qt}^{s}\left(T_{\theta_{l}}x_{ij}\right),
\vspace{-2mm}
\end{equation}
where $T_{\theta_{l}}$ is the rotation matrix with angle $\theta_{l}$ as $T_{\theta_{l}}=[\text{cos}\theta_{l}, \text{-sin}\theta_{l} ; \text{sin}\theta_{l}, \text{cos}\theta_{l}]^{T}$; $x_{ij}=[x_{i},x_{j}]^{T}=[\left(i-\sfrac{(p+1)}{2}\right)h, \left(j-\sfrac{(p+1)}{2}\right)h]^{T}$; $i=1,\ldots, p$, $j=1,\ldots, p$. ${\varphi}_{qt}^{c}\left(T_{\theta_{l}}x_{ij}\right)$ and  ${\varphi}_{qt}^{s}\left(T_{\theta_{l}}x_{ij}\right)$ are rotated Fourier bases, and their definitions are in~\eqref{eq3}. The expansion coefficients $a_{qtk}$ and $b_{qtk}$ are shared among different rotations, which reflects the rotation symmetry of artifacts. Driven by the powerful fitting ability of convolutional neural network (CNN), we can automatically learn $a_{qtk}$ and $b_{qtk}$ from training data and then flexibly extract $\C_{k}(\theta_{l})$ (see Fig.~\ref{figverf}).


\vspace{1mm}

\vspace{1mm}
\noindent \textbf{Optimization Algorithm.}
By substituting Eq.~(\ref{eq7}) into Eq.~(\ref{eq4}), we can derive:
\begin{equation}\label{eq9}
\I\odot\Y=\I\odot\X +\I\odot\left(\mC\otimes \mM\right),
\end{equation}
where $\mathcal{C}\in \mathbb{R}^{p\times p \times LK}$ and $\mathcal{M}\in \mathbb{R}^{H\times W \times LK}$ are stacked by $\C_{k}(\theta_{l}) $ and $\M_{lk} $, respectively. Clearly, given a metal-affected CT image $\Y$, we need to estimate the artifact-removed CT image $\X$ and feature map $\mM$. Note that as explained in~\eqref{eq8}, $\mC$ can be flexibly learnt from data. Then we adopt the maximum-a-posterior framework and formulate the optimization problem as:
\begin{equation}\label{eq10}
\begin{split}
\min_{{\mM,\X}}\left\|\I\odot\left(\Y\!-\!\X\!-\!\mC\!\otimes\!\mM\right)\right\|_F^{2}+\alpha f_{1}(\mM)\!+\!\beta f_{2}(\X),
\end{split}
\end{equation}
\normalsize
where $\alpha$ and $\beta$ are trade-off parameters; $\|\cdot\|_{F}$ is the Frobenius norm; $f_{1}(\cdot)$ and $f_{2}(\cdot)$ are regularizers representing the prior information of $\mM$ and $\X$, respectively. Similar to~\cite{wang2021dicdnet,wang2021indudonet}, we also adopt CNN to flexibly learn the implicit priors underlying $\mM$ and $\X$ from training datasets. The details are given in Sec.~\ref{sec:net}.

To solve the problem~\eqref{eq10}, we can directly adopt the iterative algorithm in~\cite{wang2021dicdnet}. Specifically, by utilizing the proximal gradient technique~\cite{beck2009fast} to alternatively update $\mM$ and $\X$, one can easily derive the following iterative rules as:
\begin{equation}\label{eq11}
\left\{\begin{matrix}
  \mM^{(n)} = \mbox{prox}_{\alpha\eta_{1}}\left(\mM^{(n-1)} - \eta_{1}\nabla g_{1}\left(\mM^{(n-1)}\right)  \right),\\
\X^{(n)} = \mbox{prox}_{\beta\eta_{2}}\left(\X^{(n-1)} -\eta_{2}\nabla g_{2}\left(\X^{(n-1)}\right)  \right),
\end{matrix}\right.
\end{equation}
where $\nabla g_{1} \left(\mM^{(n-1)}\right)=\mC\otimes^{T}\!\left(\I\odot\left(\mC \otimes \mM^{(n-1)}\!+\!\X^{(n-1)}\!-\!\Y\right)\right)$; $\nabla g_{2} \left(\X^{(n-1)}\right)=\I\odot\left(\mC \otimes \mM^{(n)}\!+\!\X^{(n-1)}\!-\!\Y\right)$;
$\otimes^T$ is transposed convolution; $\eta_{1}$ and $\eta_{2}$ are stepsizes; $\mbox{prox}_{\alpha\eta_1}(\cdot)$ and $\mbox{prox}_{\beta\eta_2}(\cdot)$ are proximal operators, relying on the priors $f_{1}(\cdot)$ and $f_{2}(\cdot)$, respectively, and the detailed designs are presented in Sec.~\ref{sec:net}.

\begin{figure*}[t]
  \begin{center}
     \includegraphics[width=0.9\linewidth]{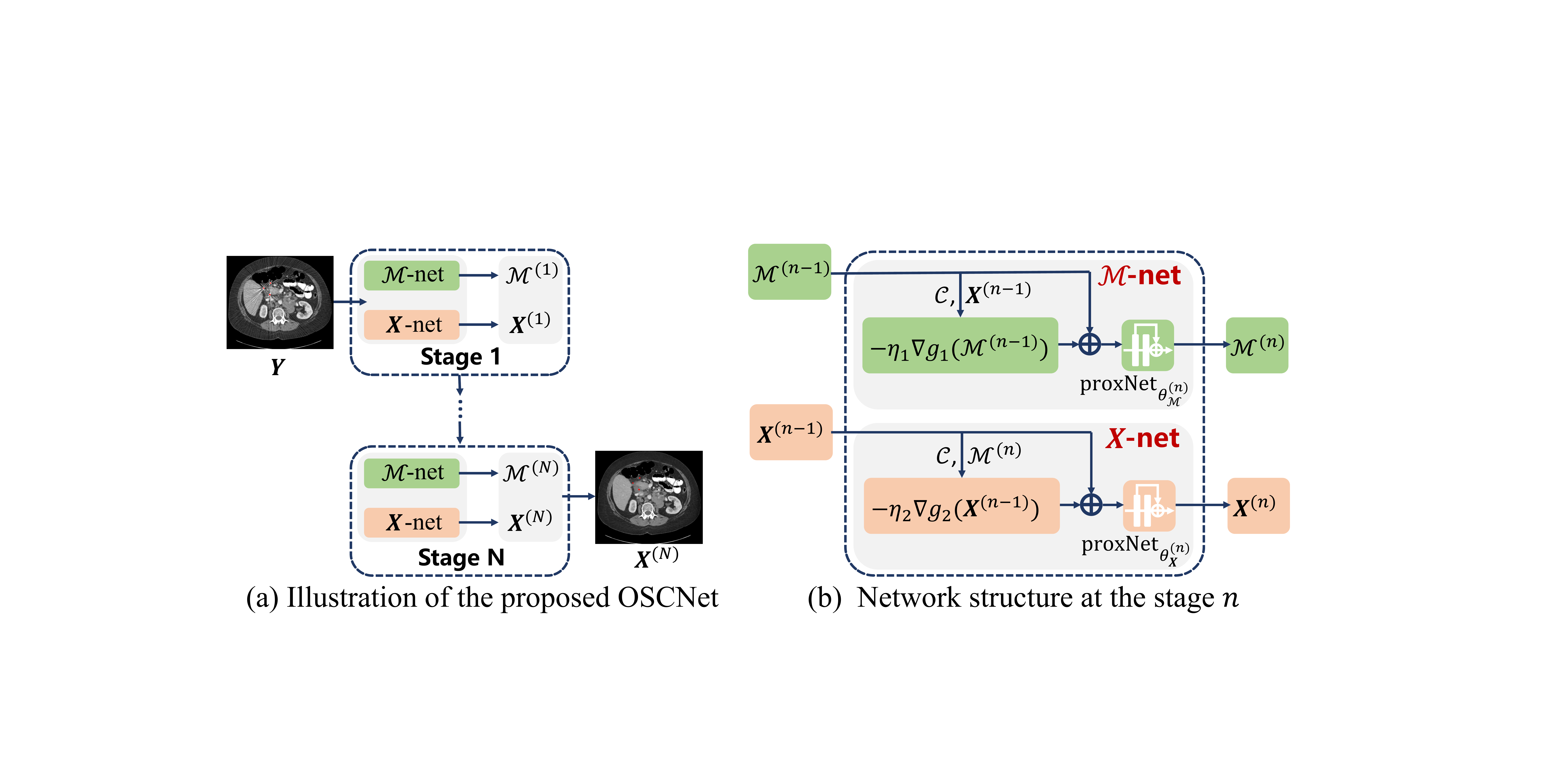}
  \end{center}
  \vspace{-6mm}
     \caption{(a) Illustration of our OSCNet. (b) At any stage $n$, the network sequentially consists of $\mM$-net and $\X$-net, which are built based on the iterative rules~\eqref{eq11}.}
  \label{fignet}
    \vspace{-2mm}
\end{figure*}

\vspace{-2mm}
\section{Network Design and Implementation Details}\label{sec:net}
Following DICDNet~\cite{wang2021dicdnet}, we can easily build an optimization-inspired deep network by sequentially unfolding every computation step in~\eqref{eq11} into the corresponding network module~\cite{wang2022adaptive}. Specifically, as shown in Fig.~\ref{fignet}, the proposed orientation-shared convolutional network (called OSCNet) contains $N$ stages, corresponding to $N$ optimization iterations. At each stage, OSCNet sequentially consists of $\mM$-net and $\X$-net for optimizing $\mM$ and $\X$, respectively. All the network parameters contain expansion coefficients $\{a_{qtk},b_{qtk}\}_{q=0,t=0,k=1}^{p-1,p-1,K}$, stepsizes $\eta_{1}$ and $\eta_{2}$, and the weights of $\mM$-net and $\X$-net as $\{\theta_\mM^{(n)},\theta_{\X}^{(n)}\}_{n=1}^{N}$, which are learnable in an end-to-end manner. At stage $n$, the network structure is:
\begin{equation}\label{eq13}
\left\{\begin{matrix}
\mM \text{-net:}~  \mM^{(n)} = \text{proxNet}_{\theta_\mM^{(n)}}\left(\mM^{(n-1)} - \eta_{1}\nabla g_{1}\left(\mM^{(n-1)}\right)  \right),\\
\X \text{-net:}~ \X^{(n)} = \text{proxNet}_{\theta_{\X}^{(n)}}\left(\X^{(n-1)} -\eta_{2}\nabla g_{2}\left(\X^{(n-1)}\right)  \right),
\end{matrix}\right.
\end{equation}
where consistent to DICDNet~\cite{wang2021dicdnet}, $\text{proxNet}_{\theta_\mM^{(n)}}(\cdot)$ and $\text{proxNet}_{\theta_{\X}^{(n)}}(\cdot)$ are ResNets with three [{\normalsize{\emph{Conv+BN+ReLU+Conv+BN+Skip Connection}}}] residual blocks, representing the proximal operators $\mbox{prox}_{\alpha\eta_1}(\cdot)$ and $\mbox{prox}_{\beta\eta_2}(\cdot)$ in~\eqref{eq11}, respectively. Besides, since the formation of artifacts is complicated, we implement a ResNet after the last stage to further refine $\X^{(N)}$~\cite{wang2021dicdnet}. It should be noted that the learnable parameter for representing filters is a common convolution layer for DICDNet but expansion coefficients for our OSCNet.

In such an optimization-inspired network design manner, our OSCNet is naturally integrated with the prior model~\eqref{eq7}. With the aid of such prior knowledge, OSCNet can achieve better model generalizability in artifact learning (see Sec.~\ref{sec:exp}). Compared to most current MAR methods, which heuristically design deep networks, the physical meanings of every module in OSCNet, correspondingly built based on the iterative rules~\eqref{eq11}, are clearer. It is worth mentioning that OSCNet can be easily built by embedding the prior model~\eqref{eq7} into the current SOTA framework~\cite{wang2021dicdnet}. This is friendly for model deployment in practices.

It is worth mentioning that as compared with DICDNet~\cite{wang2021dicdnet},  the proposed OSCNet contains novel and challenging design: 1) Eq.~(\ref{eq7}) should be the first orientation shared convolution coding model, which is hard to achieve without the filter parametrization strategy we adopt; 2) The OSC model is very suitable for encoding the rotational prior of artifacts ignored by DICDNet; 3) DICDNet is a special case of OSCNet ($L=1$, $K=32$). Such rotational design makes OSCNet outperform DICDNet obviously (See Sec.~\ref{sec:exp}); 4) The OSC model would be used as a general tool for modelling any rotational structure, which is valuable.

\vspace{1mm}
\noindent\textbf{Loss Function.}
The total objective function for training OSCNet is set as~\cite{wang2021dicdnet}:
\scriptsize
\begin{equation}\label{Loss}
\begin{split}
  \mathcal{L} \!=\!\! \!\sum_{n=0}^{N}\!\mu_{n}\I\!\odot\!\left\|\X\!-\!\X^{(n)} \right\|_F^2\!\!+\!\!\lambda_{1}\left(\!\sum_{n=0}^{N}\mu_{n}\I\!\odot\!\left\|\X\!-\!\X^{(n)} \right\|_1\!\!\right) \!\!+\!\!\lambda_{2}\left(\!\sum_{n=1}^{N}\mu_{n}\I\!\odot\!\left\|\Y\!-\!\X\!-\A^{(n)} \right\|_1\!\!\right),
\end{split}
\end{equation}
\normalsize
where $\X^{(n)}$ and $\A^{(n)}$ are the artifact-reduced image and the extracted artifact at the stage $n$, respectively; $\A^{(n)}=\mC\otimes\mM^{(n)}$; $\X$ is the ground truth CT image; $\lambda_{1}$, $\lambda_{2}$, and $\mu_{n}$ are trade-off parameters. Empirically, we set: $\lambda_{1}=\lambda_{2}=5\times 10^{-4}$, $\mu_{N}=1$, and $\mu_{n}=0.1 \,\,\,(n\neq N)$. $\X^{(0)}$ is initialized in the same manner to~\cite{wang2021dicdnet}.

\vspace{1mm}
\noindent\textbf{Training Details.} 
OSCNet is implemented using PyTorch~\cite{paszke2017automatic} and optimized on an NVIDIA Tesla V100-SMX2 GPU based on Adam optimizer with parameters ($\beta_{1}$, $\beta_{2}$)=(0.5, 0.999). The learning rate is $2\times10^{-4}$ and it is multiplied by 0.5 after every 30 training epochs. The number of total epochs is 200. The batch size is 16 and the patch size is $64 \times 64$. For a fair comparison, consistent to~\cite{wang2021dicdnet}, $N=10$, $p=9$, $L=8$, $K=4$. The mesh size $h$ in Eq.~\eqref{eq3} is $\sfrac{1}{4}$~\cite{xie2021fourier}.

\begin{figure*}[t]
  \begin{center}
     \includegraphics[width=0.82\linewidth]{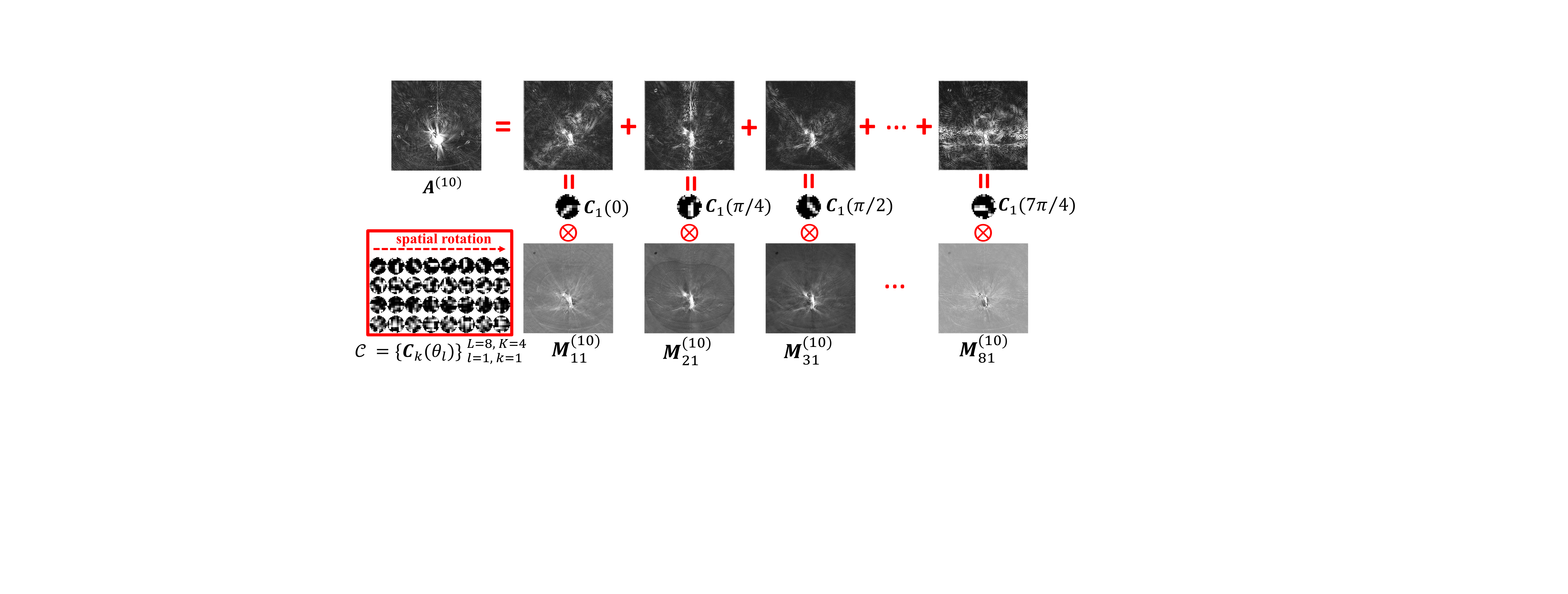}
  \end{center}
  \vspace{-6mm}
     \caption{Visualization of the model~\eqref{eq7}. $\C_{k}(\theta_{l})$, $\M_{lk}^{(10)}$, and $\A^{(10)}$ are convolution filters, feature map, and artifacts, respectively, extracted from the last stage of OSCNet. As seen, each filter in $\mC$ can adaptively fit a local structure with different rotations.}
  \label{figverf}
    \vspace{-2mm}
\end{figure*}
\begin{figure*}[t]
  \begin{center}
     \includegraphics[width=0.82\linewidth]{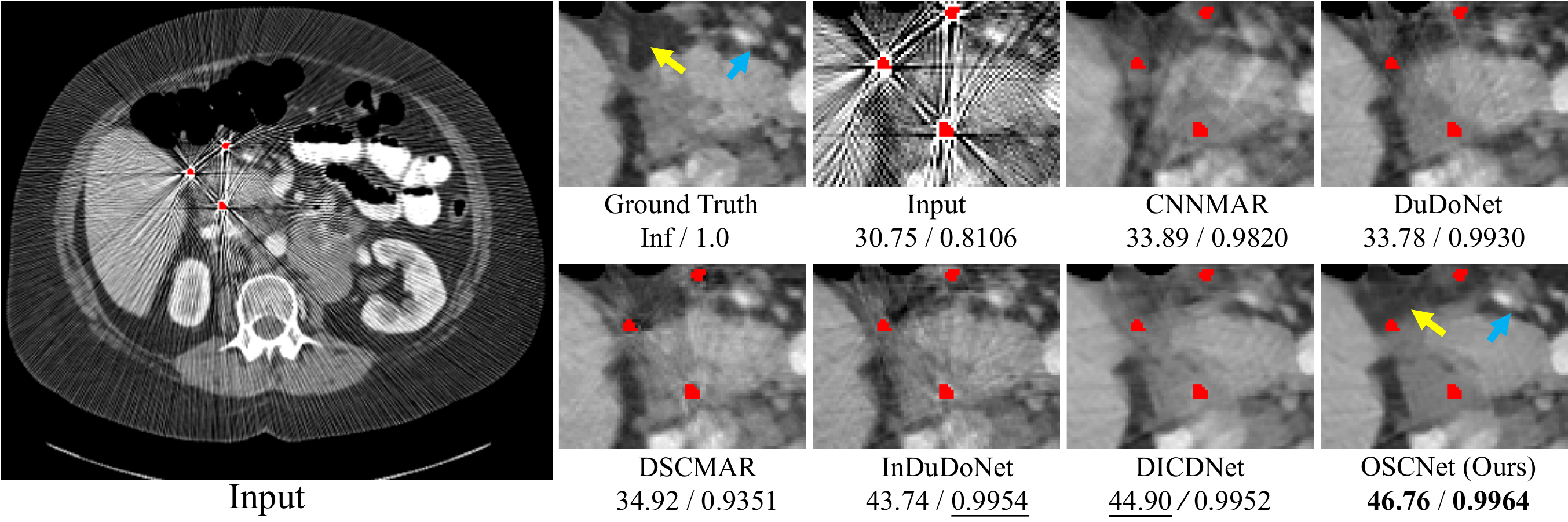}
  \end{center}
  \vspace{-6mm}
     \caption{Performance comparison on synthesized DeepLesion~\cite{yan2018deep}. PSNR/SSIM is listed below each artifact-reduced CT image. Red pixels represent metallic implants.}
  \label{figsyn}
    \vspace{-3mm}
\end{figure*}

\vspace{-1mm}
\section{Experiments}\label{sec:exp}
\noindent\textbf{Synthesized DeepLesion.}
Following~\cite{wang2021dicdnet,yu2020deep}, we can synthesize the paired $\X$ and $\Y$ for training and testing by utilizing 1,200 clean CT images from DeepLesion (mainly focusing on abdomen and thorax)~\cite{yan2018deep} and 100 simulated metallic implants from~\cite{zhang2018convolutional}. All the CT images are resized to $416\times 416$ pixels and 640 projection views are uniformly sampled in [$0^{\text{o}}$, $360^{\text{o}}$]. The sizes of 10 testing metal masks are [2061, 890, 881, 451, 254, 124, 118, 112, 53, 35] in pixels. We take adjacent masks as one group for performance evaluation.

\noindent\textbf{Synthesized Dental.}
For cross-body-site performance comparison, we collect clean dental CT images from~\cite{yu2020deep} and adopt the simulation procedure~\cite{wang2021dicdnet,yu2020deep} to synthesize paired $\X$ and $\Y$ for the evaluation of model generalization. 

\noindent\textbf{Clinical CLINIC-metal.}
Besides, we collect the public metal-affected CLINIC-metal~\cite{liu2020deep} (focusing on pelvic CT) for clinical evaluation. Consistent to~\cite{liao2019adn,wang2021dicdnet}, clinical metallic implants are roughly segmented with a threshold of 2,500 HU.

%
\begin{table}[t]
\centering
\caption{Average PSNR (dB)/SSIM on synthesized DeepLesion~\cite{yan2018deep}.}\vspace{-2mm}
\tiny
\setlength{\tabcolsep}{3.7pt}
\begin{tabular}{l|c|c|c|c|c|c}
\hline
Methods    & \multicolumn{5}{c|}{ Large Metal \quad \quad   \quad\quad  $\longrightarrow$ \quad \quad Medium Metal \quad \quad $\longrightarrow$    \quad   \quad        Small Metal}                & Average      \\
\hline
Input             &24.12/0.6761              &26.13/0.7471              &27.75/0.7659               &28.53/0.7964              &28.78/0.8076              &27.06/0.7586             \\
LI~\cite{kalender1987reduction}              &27.21/0.8920              &28.31/0.9185              &29.86/0.9464              &30.40/0.9555              &30.57/0.9608              &29.27/0.9347   \\
NMAR~\cite{meyer2010normalized}              &27.66/0.9114              &28.81/0.9373              &29.69/0.9465              &30.44/0.9591              &30.79/0.9669              &29.48/0.9442              \\
CNNMAR~\cite{zhang2018convolutional}            &28.92/0.9433  &29.89/0.9588  & 30.84/0.9706             &31.11/0.9743              &31.14/0.9752              &30.38/0.9644               \\
DuDoNet~\cite{lin2019dudonet}           & 29.87/0.9723 & 30.60/0.9786 & 31.46/0.9839  & 31.85/0.9858 & 31.91/0.9862 & 31.14/0.9814   \\
DSCMAR~\cite{yu2020deep}           & 34.04/0.9343 & 33.10/0.9362 & 33.37/0.9384  & 32.75/0.9393 & 32.77/0.9395 & 33.21/0.9375 \\
InDuDoNet~\cite{wang2021indudonet}& {36.74/0.9742} &{39.32}/{0.9893} & {41.86}/\underline{0.9944} &{44.47}/{0.9948} & {45.01}/\underline{0.9958}& {41.48}/{0.9897}\\
DICDNet~\cite{wang2021dicdnet}&\underline{37.19}/\underline{0.9853} &\underline{39.53}/\textbf{0.9908} & \underline{42.25}/{0.9941}  &\underline{44.91}/\underline{0.9953} & \underline{45.27}/\underline{0.9958} &\underline{41.83}/\underline{0.9923}
\\

OSCNet(Ours) & \textbf{37.70}/\textbf{0.9883} &\textbf{39.88}/\underline{0.9902} &\textbf{42.92}/\textbf{0.9950} &\textbf{45.04}/\textbf{0.9958} &\textbf{45.45}/\textbf{0.9962} &\textbf{42.19}/\textbf{0.9931} \\

\hline
\end{tabular}
\vspace{-0mm}
\label{tabsyn}
\end{table}
\begin{figure*}[t]
  \begin{center}
     \includegraphics[width=1\linewidth]{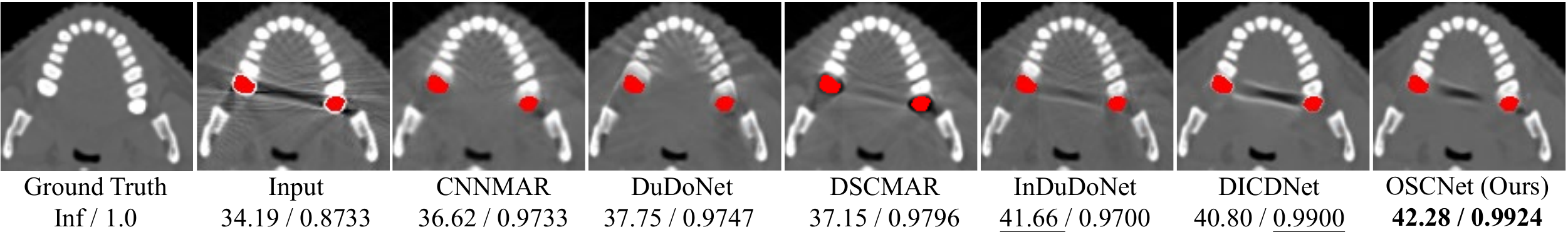}
  \end{center}
  \vspace{-7mm}
     \caption{Generalization comparison on synthesized Dental~\cite{yu2020deep}.}
  \label{figdental}
    \vspace{-1mm}
\end{figure*}
\begin{figure*}[!htp]
  \begin{center}
     \includegraphics[width=1\linewidth]{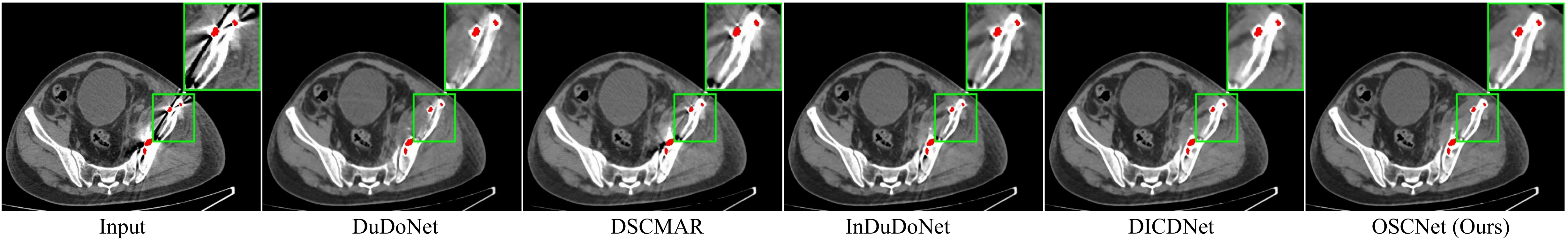}
  \end{center}
  \vspace{-7mm}
     \caption{Generalization comparison on CLINIC-metal~\cite{liu2020deep}.}
  \label{figclinic}
    \vspace{-4mm}
\end{figure*}
\noindent\textbf{Evaluation Metrics.} For synthesized data, we adopt PSNR/SSIM; For CLINIC-metal, due to the lack of ground truth, we provide visual results.

\noindent\textbf{Baselines.}
We compare with traditional LI~\cite{kalender1987reduction} and NMAR~\cite{meyer2010normalized}, DL-based CNNMAR~\cite{zhang2018convolutional}, DuDoNet~\cite{lin2019dudonet}, DSCMAR~\cite{yu2020deep}, InDuDoNet~\cite{wang2021indudonet}, and DICDNet~\cite{wang2021dicdnet}. 


\vspace{1mm}
\noindent\textbf{Model Visualization.}
Fig.~\ref{figverf} presents the convolution filter $\C_{k}(\theta_{l})$, feature map $\M_{lk}^{(10)}$, and metal artifacts $\A^{(10)}$ extracted by our OSCNet at the last stage. It is easily observed that $\C_{k}$ contains diverse patterns and is spatially rotated with angle $\frac{\pi}{4}(l-1),l=1,2,\cdots,8$. Correspondingly, the extracted artifact layer $\C_{k}(\theta_{l})\otimes \M_{lk}^{(10)}$ clearly shows rotated structures. The observation verifies the effectiveness of our OSCNet in the learning of RSS prior structures underlying the MAR task, which complies with the Fourier-series-expansion-based orientation-shared convolution prior model~\eqref{eq7}. Integrated with such prior regularization, OSCNet would recover CT images with better visual quality as shown below.


\vspace{1mm}
\noindent\textbf{Experiments on Synthesized DeepLesion.}
Fig.~\ref{figsyn} shows the artifact-reduced CT images reconstructed by all comparing methods. Clearly, our OSCNet better removes the artifacts while preserving the body tissues, especially around metallic implants. Compared to~\cite{wang2021dicdnet}, the accurate embedding of rotation prior structures can indeed benefit the identification of metal artifacts. Table~\ref{tabsyn} lists the average quantitative results on synthesized DeepLesion. As shown, our OSCNet achieves the highest PSNR score for metals with different sizes. The experimental results finely verify the effectiveness of the prior model~\eqref{eq7}. Actually, DICDNet is exactly our backbone and represents ablation study.

\vspace{1mm}
\noindent\textbf{Experiments on Synthesized Dental.} 
Fig.~\ref{figdental} compares the cross-body-site performances on synthesized Dental. 
As seen, there is still a black band in the CT image reconstructed by our method. The underlying explanation is that compared to the black-box network learning manner for artifact reduction, the explicit regularization on artifacts (\emph{i.e.}, OSC model) would weaken the flexibility of artifact extraction to some extent. Even in this case, our method still obtains the best PSNR/SSIM for fine detail fidelity and shows better model generalizability. This is mainly attributed to the more accurate regularization, which can better help network distinguish artifact layer from body tissues and guarantee the detail fidelity. This would be more meaningful for clinical applications.

\vspace{1mm}
\noindent\textbf{Experiments on CLINIC-metal.}
Fig.~\ref{figclinic} provides the reconstruction results on CLINIC-metal. As seen, our proposed OSCNet outperforms other baselines on the removal of shadings and streaking artifacts, and more accurately recovers the bone areas, which demonstrates its better clinical generalization ability.

\textbf{More experiments are provided in \emph{supplementary material}.}

\vspace{-2mm}
\section{Conclusion and Future Work}
\vspace{-2mm}
In this paper, against the metal artifact reduction task, we proposed an orientation-shared convolution model to encode the inherent rotationally symmetrical streaking prior structures of metal artifacts. By embedding such prior model into a current SOTA framework, we built an optimization-inspired deep network with clear interpretability. Comprehensive experiments on publicly available datasets have shown the rationality and effectiveness of our proposed method. However, there still exists some potential limitation, such as, metal segmentation might make tissues be wrongly regarded as metals and most MAR methods as well as our OSCNet would fail to recover image details. How to accomplish the joint optimization of automated metal localization and accurate metal artifact reduction deserves further exploration in the future.

\vspace{-4mm}
\subsubsection{Acknowledgements.}
This work was founded by the China NSFC project under contract U21A6005, the Macao Science and Technology Development Fund under Grant 061/2020/A2, the major key project of PCL under contract PCL2021A12, the Key-Area Research and Development Program of Guangdong Province, China (No. 2018B010111001), National Key R\&D Program of China (2018YFC2000702), the Scientific and Technical Innovation 2030-``New Generation Artificial Intelligence'' Project (No. 2020AAA0104100).
%
%
\bibliographystyle{splncs04}
\bibliography{mybib}

\begin{thebibliography}{10}
\providecommand{\url}[1]{\texttt{#1}}
\providecommand{\urlprefix}{URL }
\providecommand{\doi}[1]{https://doi.org/#1}

\bibitem{beck2009fast}
Beck, A., Teboulle, M.: A fast iterative shrinkage-thresholding algorithm for
  linear inverse problems. SIAM Journal on Imaging Sciences  \textbf{2}(1),
  183--202 (2009)

\bibitem{cooley1969fast}
Cooley, J.W., Lewis, P.A., Welch, P.D.: The fast {F}ourier transform and its
  applications. IEEE Transactions on Education  \textbf{12}(1),  27--34 (1969)

\bibitem{ghani2019fast}
Ghani, M.U., Karl, W.C.: Fast enhanced {C}{T} metal artifact reduction using
  data domain deep learning. IEEE {T}ransactions on {C}omputational {I}maging
  \textbf{6},  181--193 (2019)

\bibitem{gjesteby2018deep}
Gjesteby, L., Shan, H., Yang, Q., Xi, Y., Claus, B., Jin, Y., De~Man, B., Wang,
  G.: Deep neural network for {C}{T} metal artifact reduction with a perceptual
  loss function. In: Proceedings of the Fifth International Conference on Image
  Formation in X-ray Computed Tomography (2018)

\bibitem{gjesteby2017deep}
Gjesteby, L., Yang, Q., Xi, Y., Zhou, Y., Zhang, J., Wang, G.: Deep learning
  methods to guide {C}{T} image reconstruction and reduce metal artifacts. In:
  Proceedings of the SPIE Conference on Medical Imaging: Physics of Medical
  Imaging. vol. 10132, p. 101322W (2017)

\bibitem{huang2018metal}
Huang, X., Wang, J., Tang, F., Zhong, T., Zhang, Y.: Metal artifact reduction
  on cervical {C}{T} images by deep residual learning. Biomedical {E}ngineering
  {O}nline  \textbf{17}(1),  1--15 (2018)

\bibitem{kalender1987reduction}
Kalender, W.A., Hebel, R., Ebersberger, J.: Reduction of {C}{T} artifacts
  caused by metallic implants. Radiology  \textbf{164}(2),  576--577 (1987)

\bibitem{liao2019generative}
Liao, H., Lin, W.A., Huo, Z., Vogelsang, L., Sehnert, W.J., Zhou, S.K., Luo,
  J.: Generative mask pyramid network for {C}{T}/{C}{B}{C}{T} metal artifact
  reduction with joint projection-sinogram correction. In: International
  Conference on Medical Image Computing and Computer Assisted Intervention. pp.
  77--85 (2019)

\bibitem{liao2019adn}
Liao, H., Lin, W.A., Zhou, S.K., Luo, J.: {A}{D}{N}: Artifact disentanglement
  network for unsupervised metal artifact reduction. IEEE Transactions on
  Medical Imaging  \textbf{39}(3),  634--643 (2019)

\bibitem{lin2019dudonet}
Lin, W.A., Liao, H., Peng, C., Sun, X., Zhang, J., Luo, J., Chellappa, R.,
  Zhou, S.K.: Du{D}o{N}et: Dual domain network for {C}{T} metal artifact
  reduction. In: Proceedings of the IEEE/CVF Conference on Computer Vision and
  Pattern Recognition. pp. 10512--10521 (2019)

\bibitem{liu2020deep}
Liu, P., Han, H., Du, Y., Zhu, H., Li, Y., Gu, F., Xiao, H., Li, J., Zhao, C.,
  Xiao, L., et~al.: Deep learning to segment pelvic bones: large-scale {CT}
  datasets and baseline models. International Journal of Computer Assisted
  Radiology and Surgery  \textbf{16}(5),  749--756 (2021)

\bibitem{lyu2020dudonet++}
Lyu, Y., Lin, W.A., Liao, H., Lu, J., Zhou, S.K.: Encoding metal mask
  projection for metal artifact reduction in computed tomography. In:
  International {C}onference on {M}edical {I}mage {C}omputing and {C}omputer
  {A}ssisted {I}ntervention. pp. 147--157 (2020)

\bibitem{meyer2010normalized}
Meyer, E., Raupach, R., Lell, M., Schmidt, B., Kachelrie{\ss}, M.: Normalized
  metal artifact reduction ({N}{M}{A}{R}) in computed tomography. Medical
  Physics  \textbf{37}(10),  5482--5493 (2010)

\bibitem{park2018ct}
Park, H.S., Lee, S.M., Kim, H.P., Seo, J.K., Chung, Y.E.: {C}{T}
  sinogram-consistency learning for metal-induced beam hardening correction.
  Medical Physics  \textbf{45}(12),  5376--5384 (2018)

\bibitem{paszke2017automatic}
Paszke, A., Gross, S., Chintala, S., Chanan, G., Yang, E., DeVito, Z., Lin, Z.,
  Desmaison, A., Antiga, L., Lerer, A.: Automatic differentiation in {PyTorch}.
  In: {A}dvances in {N}eural {I}nformation {P}rocessing {S}ystems {W}orkshop
  (2017)

\bibitem{wang2021dicdnet}
Wang, H., Li, Y., He, N., Ma, K., Meng, D., Zheng, Y.: {DICDNet: D}eep
  interpretable convolutional dictionary network for metal artifact reduction
  in {CT} images. IEEE Transactions on Medical Imaging  (2021)

\bibitem{wang2022adaptive}
Wang, H., Li, Y., Meng, D., Zheng, Y.: Adaptive convolutional dictionary
  network for ct metal artifact reduction. arXiv preprint arXiv:2205.07471
  (2022)

\bibitem{wang2021indudonet}
Wang, H., Li, Y., Zhang, H., Chen, J., Ma, K., Meng, D., Zheng, Y.:
  {InDuDoNet}: {A}n interpretable dual domain network for {CT} metal artifact
  reduction. In: International Conference on Medical Image Computing and
  Computer Assisted Intervention. pp. 107--118 (2021)

\bibitem{wang2021indudonet+}
Wang, H., Li, Y., Zhang, H., Meng, D., Zheng, Y.: {InDuDoNet+}: A model-driven
  interpretable dual domain network for metal artifact reduction in {CT}
  images. arXiv preprint arXiv:2112.12660  (2021)

\bibitem{wang2018conditional}
Wang, J., Zhao, Y., Noble, J.H., Dawant, B.M.: Conditional generative
  adversarial networks for metal artifact reduction in {C}{T} images of the
  ear. In: International Conference on Medical Image Computing and Computer
  Assisted Intervention. pp. 3--11 (2018)

\bibitem{wang2021dualclean}
Wang, T., Xia, W., Huang, Y., Sun, H., Liu, Y., Chen, H., Zhou, J., Zhang, Y.:
  Dual-domain adaptive-scaling non-local network for {CT} metal artifact
  reduction. In: International Conference on Medical Image Computing and
  Computer Assisted Intervention. pp. 243--253. Springer (2021)

\bibitem{xie2021fourier}
Xie, Q., Zhao, Q., Xu, Z., Meng, D.: Fourier series expansion based filter
  parametrization for equivariant convolutions. arXiv preprint arXiv:2107.14519
   (2021)

\bibitem{yan2018deep}
Yan, K., Wang, X., Lu, L., Zhang, L., Harrison, A.P., Bagheri, M., Summers,
  R.M.: Deep lesion graphs in the wild: Relationship learning and organization
  of significant radiology image findings in a diverse large-scale lesion
  database. In: Proceedings of the IEEE Conference on Computer Vision and
  Pattern Recognition. pp. 9261--9270 (2018)

\bibitem{yu2020deep}
Yu, L., Zhang, Z., Li, X., Xing, L.: Deep sinogram completion with image prior
  for metal artifact reduction in {C}{T} images. IEEE Transactions on Medical
  Imaging  \textbf{40}(1),  228--238 (2020)

\bibitem{zhang2018convolutional}
Zhang, Y., Yu, H.: Convolutional neural network based metal artifact reduction
  in {X}-ray computed tomography. IEEE Transactions on Medical Imaging
  \textbf{37}(6),  1370--1381 (2018)

\bibitem{zhou2022dudodr}
Zhou, B., Chen, X., Zhou, S.K., Duncan, J.S., Liu, C.: {DuDoDR-N}et:
  Dual-domain data consistent recurrent network for simultaneous sparse view
  and metal artifact reduction in computed tomography. Medical Image Analysis
  \textbf{75},  102289 (2022)

\end{thebibliography}
\end{document}